\documentclass[12pt]{article}
\usepackage[centertags]{amsmath}
\usepackage{epsfig}
\usepackage{doublespace}
\usepackage{amssymb}
\usepackage{latexsym}
\usepackage{vmargin}
\setpapersize{A4}
\setmarginsrb{35mm}{20mm}{20mm}{20mm}{0pt}{0mm}{0pt}{5mm}
\allowdisplaybreaks

\newcommand{\ga}{\ensuremath{\gamma}}
\newcommand{\be}{\ensuremath{\beta}}
\newcommand{\al}{\ensuremath{\alpha}}

\newcommand{\p}{\ensuremath{\partial}}

\newcommand{\si}{\ensuremath{\sigma}}
\newcommand{\sfrac}[2]{\ensuremath{{\scriptstyle \frac{#1}{#2}}}}

\begin{document}

\begin{spacing}{1.0}

\setlength{\abovedisplayskip}{10pt plus 3pt minus 9pt}
\setlength{\belowdisplayskip}{10pt plus 3pt minus 9pt}
\setlength{\abovedisplayshortskip}{0pt plus 3pt}
\setlength{\belowdisplayshortskip}{5pt plus 3pt minus 4pt}

\title{\bf Perfect-fluid cylinders and walls---sources for the Levi--Civita 
space--time}

\author{Thomas G. Philbin\thanks{E-mail: tphilbin@maths.tcd.ie.} \\
 \small \it  School of Mathematics, Trinity College, Dublin 2, Ireland}

\date{}

\maketitle

\begin{abstract}

The diagonal metric tensor whose components are functions of one spatial 
coordinate is considered. Einstein's field equations for a perfect-fluid 
source are reduced to quadratures once a generating function, equal to the 
product of two of the metric components, is chosen. The solutions are either 
static fluid cylinders or walls depending on whether or not one of the spatial 
coordinates is periodic. Cylinder  and wall sources are generated and matched 
to the vacuum (Levi--Civita) space--time. A match to a cylinder source is 
achieved for $-\frac{1}{2}<\si<\frac{1}{2}$, where $\si$ is the mass per unit 
length in the Newtonian limit $\si\to 0$, and a match to a wall source is 
possible for $|\si|>\frac{1}{2}$, this case being without a Newtonian limit; 
the positive (negative) values of $\si$ correspond to a positive (negative) 
fluid density. The range of $\si$ for which a source has previously been 
matched to the Levi--Civita metric is $0\leq\si<\frac{1}{2}$ for a cylinder 
source.
\end{abstract}

\end{spacing}
\begin{spacing}{1.5}

\section{Introduction}
Although a large number of vacuum solutions of Einstein's equations are known, 
the physical interpretation of many (if not most) of them remains unsettled
(see, e.g.,~\cite{Bonnor92}). As stressed by Bonnor~\cite{Bonnor92}, the key to
physical interpretation is to ascertain the nature of the sources which produce 
these vacuum space--times; even in black-hole solutions, where no matter source
is needed, an understanding of how a matter distribution gives rise to the 
black-hole space--time is necessary to judge the physical significance (or lack
of it) of the complete analytic extensions of these solutions. 

As to how we gain an understanding of the sources, there is really no 
substitute for constructing an interior solution for a matter distribution 
which matches to the vacuum space--time in question. The coordinate freedom of 
general relativity can make attempts to discover the nature of the source for 
a vacuum field a hazardous affair: in the vacuum space--time the source appears
as a singularity in the curvature which may look quite different after a 
coordinate transformation~\cite{Bonnor91}. If we have an interior solution 
however such coordinate transformations will invariably \emph{introduce} a 
singularity in the interior and so must be rejected. Although attempts are made
to deduce properties of sources by analysis of the vacuum space--times which 
represent the exterior fields, this approach can never be conclusive; only a 
complete interior and exterior solution will afford confidence in the analysis.

In this paper we construct sources for the vacuum space--time described by a
diagonal metric depending on one spatial coordinate; this vacuum solution,
found by Tullio Levi--Civita~\cite{Levi}, we shall write as
\begin{equation}  \label{Levi}
ds^2=A^2(r-k)^{8\si^2-4\si}(dr^2+dz^2)+B^2(r-k)^{2-4\si}d\phi^2
                        -C^2(r-k)^{4\si}dt^2,
\end{equation}
where $\si$, $k$, $A$, $B$ and $C$ are arbitrary constants.
(In fact, by rescaling, (\ref{Levi}) can be cast in the form
\[
ds^2=r^{8\si^2-4\si}(dr^2+dz^2)+D^2r^{2-4\si}d\phi^2-r^{4\si}dt^2,
\]
but the constant $D$ cannot be removed if $\phi$ is to be a periodic coordinate
with period $2\pi$~\cite{Marder}. Thus the Levi--Civita metric has two
parameters when $\phi$ is periodic. The form (\ref{Levi}) is needed in 
matching to the interior solutions.) This space--time has in general a 
curvature singularity at $r=k$ and is flat in the limit $r\to\infty$; the 
Riemann tensor vanishes everywhere only for $\si=0$ and $\si=\frac{1}{2}$. 

A test particle at rest in the coordinate system of the metric (\ref{Levi}) 
will experience a proper acceleration
\begin {equation}   \label{accel}
\ddot{r}=-\frac{2\si}{(r-k)^{1+8\si^2-4\si}}.
\end{equation}
For small $\si$ this is approximately
\[
\ddot{r}=-\frac{2\si}{r-k},
\]
which is of the same form as the Newtonian expression for the acceleration of a
particle a distance $r-k$ from a line mass of mass per unit length $\si$. This 
well-defined Newtonian limit is the reason why the Levi-Civita space--time is
usually said to represent the field outside an infinitely long static 
cylinder---thus the $\phi$-coordinate is taken to be periodic---and cylinder 
sources have indeed been found for this 
space--time~\cite{Bonnor79,Lathrop,Bonnor92b,daSilva}. The sources found are of
two types: static dust cylinders (composed of equal amounts of dust rotating in
opposite senses around the axis to produce zero net angular momentum) which
were shown to match to the Levi--Civita space--time for values of $\si$ in the
range $0\leq\si<\frac{1}{4}$~\cite{Lathrop,daSilva}, and perfect-fluid 
cylinders, one example of which could be matched to the Levi--Civita exterior
for $0\leq\si<\frac{1}{4}$~\cite{Bonnor79} while the other example was valid
for $0\leq\si<\frac{1}{2}$~\cite{Bonnor92b}. These results are consistent with 
the interpretation of the Levi-Civita space--time as the 
relativistic field outside a line mass, at least for $\si$ in the range 
$0\leq\si<\frac{1}{2}$. 
\footnote{It is worth mentioning that for $\si\geq\frac{1}{4}$ the Levi--Civita
space--time does not contain timelike circular geodesics~\cite{Gautreau} (when
$\si=\frac{1}{4}$ the circular geodesics are null, when $\si>\frac{1}{4}$ they
are spacelike). The explanation, suggested by the Newtonian case, that this is 
due to the gravitational attraction becoming strong enough to prevent any 
particle (including light) orbiting circularly~\cite{Ray} has been contrasted 
with the fact that a study of the curvature invariants suggests that the field 
gets weaker as $\si$ increases from $\frac{1}{4}$ to $\frac{1}{2}$ (where the 
metric is flat but accelerated)~\cite{Bonnor91,daSilva}. However, although it 
sounds strange, what matters here is not the field (i.e.\ the curvature) but 
the acceleration experienced by the test particle towards the source: this does
not vanish at $\si=\frac{1}{2}$ although the field does, and indeed the proper 
acceleration towards the origin experienced by a test particle at rest in the 
coordinate system of (\ref{Levi}) (or of (\ref{Levigauss}) below) 
\emph{increases} as $\si$ goes from $\frac{1}{4}$ to $\frac{1}{2}$. Hence it is
quite reasonable to conclude that the cylinder sources producing $\si$ in the 
range $\frac{1}{4}<\si<\frac{1}{2}$ have a gravitational attraction large 
enough to disallow circular orbits.}

Nevertheless, we can just as easily take all the coordinates in (\ref{Levi}) to
be Cartesian and the supposition then is that the Levi--Civita space--time 
represents the field outside an infinite static wall; although this possibility
has been noticed before~\cite{Aichelburg} an example of such a wall source has
not been constructed. Indeed, the interpretation of $\phi$ as a Cartesian 
coordinate appears more tenable for certain values of $\si$. For example, when 
$\si=\frac{1}{2}$ the metric (\ref{Levi}) describes flat space--time in the 
local coordinate system of an observer undergoing constant acceleration in the 
$r$-direction~\cite{Misner}, and when $\si=-\frac{1}{2}$ the metric, in 
addition to the three Killing vectors $\p_t$, $\p_\phi$ and 
$\p_z$, admits a fourth Killing vector $\phi\p_z-z\p_\phi$ which identifies it 
as Taub's plane-symmetric metric~\cite{Taub} (though in different coordinates);
these two metrics are the favourite proposals for the exterior field of a plane
mass (e.g.~\cite{Schucking,Novotny,Sassi,Bedran}). There is thus a rich 
structure to the vacuum metric (\ref{Levi}) (for more detail, 
see~\cite{Bonnor91}) and we take the view that its physical significance can 
only be meaningfully explored by constructing sources which generate 
(\ref{Levi}) for a wide range of $\si$.

We shall construct both cylinder and wall sources for the Levi--Civita
space--time (\ref{Levi}), composed of perfect fluid. We work with the metric 
form
\begin{equation}  \label{synge}
ds^2=\al^2(r)(dr^2+dz^2)+\be^2(r)d\phi^2-\ga^2(r)dt^2,
\end{equation} 
the vacuum solution for which is given by (\ref{Levi}). If $\phi$ is a periodic
coordinate then this metric is cylindrically symmetric and static; if all the 
coordinates are Cartesian then the space--time is homogeneous on the
spacelike $(z,\phi)$-planes but not (in general) isotropic on those 
planes---we therefore call this latter type \emph{plane-homogeneous} 
space--times and metric (\ref{synge}) is then static and plane-homogeneous.
\footnote{The term plane-symmetric has already been used to denote space--times
which are homogeneous \emph{and} isotropic on spacelike planes~\cite{Taub},
such as Taub's metric; in order to be plane-symmetric, (\ref{synge}) must have
a third spacelike Killing vector $\phi\p_z-z\p_\phi$.
Plane-homogeneous solutions have been considered in the literature, though not
by this name, the usual practice being simply to state that there is a $G_2$ 
on $S_2$ or that there are two spacelike Killing vectors 
(see~\cite{Ruiz, Wang} and references therein); the similarity to plane
symmetry and some resulting confusion has been noted~\cite{Wang}. Li and
Liang~\cite{Li} have introduced a title for this symmetry; they found 
electrovac solutions where the metric is plane-symmetric but the 
electromagnetic field inherits only the two translational symmetries and not 
the rotational one and referred to these electromagnetic 
fields as having \emph{semi-plane symmetry}. We prefer the term 
``plane-homogeneous'' because it gives a clearer idea of the symmetry involved 
than ``semi-plane-symmetric''.}
In Sec.~2 the field equations for (\ref{synge}) with a perfect-fluid source are
reduced to quadratures once the function $\be\ga$ is chosen; the difference 
between cylindrically symmetric solutions and plane-homogeneous solutions 
emerges in the behaviour of the metric tensor at $r=0$ through the demand that 
the geometry be regular there. In Sec~3 we use this scheme to generate 
cylindrically symmetric and  plane-homogeneous solutions with a boundary (that 
is, solutions in which the pressure falls to zero at a finite value of $r$) 
which are then matched to the exterior (Levi--Civita) space--time; a cylinder 
source can be matched for $\si$ in the range $-\frac{1}{2}<\si<\frac{1}{2}$, 
and a wall source for $|\si|>\frac{1}{2}$, where the positive (negative) values
of $\si$ correspond to a positive (negative) density for the fluid. These 
results are discussed in Sec.~4.

\section{The field equations}
Einstein's equations for the metric (\ref{synge}) in geometrical units 
($c=G=1$) with a perfect fluid energy--momentum tensor 
($T^a_{\ b}=\text{diag}(p,p,p,-\rho)$) are
\begin{gather}
\frac{\al'}{\al}\left(\frac{\be'}{\be}+\frac{\ga'}{\ga}\right)
  +\frac{\be'}{\be}\frac{\ga'}{\ga}=8\pi\al^2p,  \label{fe1} \\[10pt]
-\frac{\al'}{\al}\left(\frac{\be'}{\be}
          +\frac{\ga'}{\ga}\right)+\frac{\be'}{\be}\frac{\ga'}{\ga}
          +\frac{\be''}{\be}+\frac{\ga''}{\ga}=8\pi\al^2p, \label{fe2} \\[10pt]
\left(\frac{\al'}{\al}\right)'+\frac{\ga''}{\ga}=8\pi\al^2p,\label{fe3}\\[10pt]
\left(\frac{\al'}{\al}\right)'+\frac{\be''}{\be}=-8\pi\al^2\rho \label{fe4}.  
\end{gather}
The solutions to these equations may describe either cylindrically symmetric or
plane-homogeneous space--times; the essential difference between the two types
of solution manifests itself in the boundary conditions for the equations. 

If the metric (\ref{synge}) is cylindrically symmetric with axis at $r=0$ then 
the regularity condition~\cite{Kramer}
\begin{equation}         \label{regcon}
\frac{X_{,a}X^{,a}}{4X}\to 1 \quad \text{as}\ r\to 0, 
                       \qquad X=g(\p_\phi,\p_\phi),
\end{equation}
gives
\begin{equation}   \label{beta1}
\frac{{\be'}^2}{\al^2}\to 1 \qquad \text{as $r\to 0$}.
\end{equation}
In addition the requirement (indeed, the definition) of the axis, that 
$g(\p_\phi,\p_\phi)=0$ there, gives
\begin{equation}   \label{beta2}
\be(0)=0.
\end{equation}
To simplify the discussion, we scale the coordinates so that the metric 
approaches flat space--time in cylindrical coordinates as $r\to 0$, i.e.
\begin{equation}         \label{flatcyl}
ds=dr^2+dz^2+r^2d\phi^2-dt^2 \quad \text{as}\ r\to 0.
\end{equation}
It therefore follows from (\ref{beta1}) and (\ref{beta2}) that
\begin{equation}  \label{elflat}
\be'(0)=1.
\end{equation}
In addition, it is reasonably clear from (\ref{beta2}), (\ref{elflat}) and the 
field equations (\ref{fe1})--(\ref{fe4}) that $\al'$, $\ga'$ and $\be''$ must 
also vanish at $r=0$ at least as quickly as $\be$ or else some terms in the 
Einstein tensor components diverge. In fact, Lake and Musgrave~\cite{Lake} have 
taken the set of fourteen independent second-order curvature invariants found  
by Carminati and McLenaghan~\cite{Carminati} and worked out the necessary and 
sufficient conditions for them to be finite at the origin of certain static 
space--times; for the cylindrically symmetric case the conditions are, in our 
coordinate system, those deduced above, namely
\begin{equation}   \label{regcons}
\al'(0)=0, \qquad \ga'(0)=0, \qquad \be(0)=0, \qquad \be'(0)=1, \qquad
 \be''(0)=0.
\end{equation}
Equations (\ref{regcons}) are, then, the necessary and sufficient conditions
for the metric (\ref{synge}) to describe a cylindrically symmetric space--time
which is regular on the axis. 
When the metric (\ref{synge}) describes a plane-homogeneous space--time 
however, $\be$ and the rest of the metric components must be non-zero constants
at $r=0$ and with rescaling the metric can be written
\begin{equation}         \label{flatplane}
ds^2=dr^2+dz^2+d\phi^2-dt^2 \qquad \text{at $r=0$},
\end{equation}
the coordinate $\phi$ now being a Cartesian coordinate.
It is this difference in the behaviour of the metric components at $r=0$ that
distinguishes the cylindrically symmetric solutions from the plane-homogeneous 
ones.

The content of the conservation equation $T^{ab}_{\ \ ;b}=0$ comes from the 
component orthogonal to the fluid four-velocity, 
i.e.\ $(g^{ab}+u^au^b)T_{b\ ;c}^{\ c}=0$, which yields
\begin{equation}   \label{coneqn}
\frac{\ga'}{\ga}=-\frac{p'}{p+\rho},
\end{equation}
and this relation can in fact be derived from the field equations 
(\ref{fe1})--(\ref{fe4}). Thus we have four independent equations for the five 
unknown functions $\al$, $\be$, $\ga$, $p$ and $\rho$; therefore once one 
further relation is imposed---be it an equation of state for the fluid, an 
explicit form for one of the unknowns, or any other independent 
equation---everything is determined up to arbitrary constants. What we shall 
now show is that if this additional relation is an explicit form for the 
quantity $\be\ga$, the metric components may be found by mere integration.

Adding (\ref{fe1}) and (\ref{fe2}) gives
\begin{equation}   \label{pressure}
\frac{(\be\ga)''}{\be\ga}=16\pi\al^2p.
\end{equation}
Subtracting (\ref{fe2}) from (\ref{fe3}) gives
\[
\left(\frac{\al'}{\al}\right)'+\frac{\al'}{\al}\left(\frac{\be'}{\be}
 +\frac{\ga'}{\ga}\right)-\frac{\be''}{\be}-\frac{\be'\ga'}{\be\ga}=0,
\]
which may be written
\[
\frac{1}{\be\ga}\frac{d\ }{dr}\left(\frac{\al'}{\al}\be\ga\right)
   -\frac{1}{\be\ga}\frac{d\ }{dr}\left(\be'\ga\right)=0.
\]
Multiplying across by $\be\ga$ and integrating gives
\[
\frac{\al'}{\al}\be\ga-\be'\ga=-c, \qquad \text{$c$\/ a constant}.
\]
For a cylindrically symmetric solution we must have, from 
(\ref{beta2})--(\ref{regcons}), $\be(0)=\al'(0)=0$ and $\be'(0)=\ga(0)=1$, so 
that $c=1$; $c$ remains free if the solution is plane-homogeneous.  We write 
the last formula as
\begin{equation}   \label{al}
\frac{\al'}{\al}=\frac{\be'}{\be}-\frac{c}{\be\ga}.
\end{equation}
Subtracting (\ref{fe2}) from (\ref{fe1}) we get
\[
\frac{\al'}{\al}\left(\frac{\be'}{\be}+\frac{\ga'}{\ga}\right)
  -\frac{1}{2}\frac{\be''}{\be}-\frac{1}{2}\frac{\ga''}{\ga}=0
\]
and substituting for $\frac{\al'}{\al}$ from (\ref{al}) this becomes
\begin{align*}
\left(\frac{\be'}{\be}\right)^2+\frac{\be'\ga'}{\be\ga}
  -c\frac{(\be\ga)'}{(\be\ga)^2}-\frac{1}{2}\frac{\be''}{\be}
  -\frac{1}{2}\frac{\ga''}{\ga}&=0,   \\
\intertext{or}
\left(\frac{\be'}{\be}\right)^2+2\frac{\be'\ga'}{\be\ga}
  -c\frac{(\be\ga)'}{(\be\ga)^2}-\frac{1}{2}\frac{(\be\ga)''}{\be\ga}&=0.
\end{align*}
Now 
\begin{equation}  \label{be}
\frac{\ga'}{\ga}=\frac{(\be\ga)'}{\be\ga}-\frac{\be'}{\be},
\end{equation}
so the previous equation becomes
\[
\left(\frac{\be'}{\be}\right)^2-2\frac{\be'}{\be}\frac{(\be\ga)'}{\be\ga}
  +c\frac{(\be\ga)'}{(\be\ga)^2}+\frac{1}{2}\frac{(\be\ga)''}{\be\ga}=0.
\]
This is a quadratic for $\frac{\be'}{\be}$ giving
\[
\frac{\be'}{\be}=\frac{(\be\ga)'}{\be\ga}
  \mp\sqrt{\left(\frac{(\be\ga)'}{\be\ga}\right)^2
  -c\frac{(\be\ga)'}{(\be\ga)^2}-\frac{1}{2}\frac{(\be\ga)''}{\be\ga}},
\]
or 
\begin{equation}  \label{ga}
\frac{\ga'}{\ga}=\pm\sqrt{\left(\frac{(\be\ga)'}{\be\ga}\right)^2
  -c\frac{(\be\ga)'}{(\be\ga)^2}-\frac{1}{2}\frac{(\be\ga)''}{\be\ga}}.
\end{equation}

The procedure for generating solutions is now clear: choose some function of
$r$ as $\be\ga$; then $\ga$ is found by integrating (\ref{ga}), $\be$ is
 given by $\frac{\be\ga}{\ga}$, and $\al$ is found by integrating (\ref{al}). 
The choice of $\be\ga$ determines the type of solution which will result: for 
a cylindrically symmetric solution the function $\be\ga$ must satisfy, from 
(\ref{regcons}),
\begin{equation}  \label{begacons}
(\be\ga)(0)=0, \qquad (\be\ga)'(0)=1, \qquad (\be\ga)''(0)=0,
\end{equation}
and, as explained above,  in the previous equations we put $c=1$; for a 
plane-homo\-ge\-neous solution we must have, from (\ref{flatplane}), 
$(\be\ga)(0)=1$, since the coordinates are now Cartesian, and the constant $c$ 
remains arbitrary. By first substituting the chosen $\be\ga$ into 
(\ref{pressure}) we can immediately obtain important physical information
before deciding whether or not to proceed with generating the solution; for
since $\al^2>0$ everywhere, (\ref{pressure}) shows whether the pressure is 
positive or negative and whether it reaches zero at finite $r$, thus 
allowing this value of $r$ to be taken as a boundary and a vacuum solution to 
be joined on. However the behaviour of the density $\rho$ can only be judged 
after $\frac{\ga'}{\ga}$, $\frac{\be'}{\be}$ and $\frac{\al'}{\al}$ have been 
derived and substituted into (\ref{fe4}).

Finally we show that plane-symmetric solutions are also produced by this 
procedure. We have remarked that the plane-homogeneous metric obtaining when 
$\phi$ is Cartesian in (\ref{synge}) is distinguished from the plane-symmetric 
metric by lacking a third spacelike Killing vector $\phi\p_z-z\p_\phi$. Suppose
now that (\ref{synge}) has this Killing vector; then the Killing equation 
${\displaystyle \pounds_{\phi\p_z-z\p_\phi}\,g=0}$ gives
\[
\al^2=\be^2.
\]
Now this is implemented by putting $c=0$ in (\ref{al}); hence, after generating
a plane-homogeneous solution it is simply a matter of setting $c=0$ and a 
plane-symmetric solution results.

Prior to this the only schemes for generating perfect-fluid solutions of the
form (\ref{synge}) appear to be those of Evans~\cite{Evans} and 
Kramer~\cite{Kramerb}. Starting with different coordinate systems from that
used here these authors also reduced the problem to the choosing of a 
generating function, after which everything else is determined. In these 
formulations however one must either solve a second-order linear homogeneous 
differential equation~\cite{Evans} or a pair of coupled first-order 
equations~\cite{Kramerb}. Our method has the advantage that the entire solution
is reduced to quadratures once the generating function is chosen---there are no
differential equations to solve; in addition, our method allows the immediate 
assessment of important physical information (the pressure), a feature lacking
in the aforementioned schemes.

\section{Solutions with a boundary}
These solutions are sources for the Levi--Civita space--time (\ref{Levi}), the 
physical interpretation of which will be accordingly illuminated. In this 
regard we shall be considering, in addition to solutions which are physically 
acceptable, solutions with negative mass so as clarify when the Levi--Civita 
metric represents a negative-mass source. 

The only static perfect-fluid cylinders with boundary to be found in the 
literature are those of Evans~\cite{Evans} and 
Davidson~\cite{Davidson90,Davidson90b} and we give for the first time a 
plane-homogeneous solution representing an infinite wall of perfect fluid with 
a boundary.

The boundary occurs when the pressure of the fluid vanishes at some finite 
value of $r$, $r=s$ say. The matching condition between interior and exterior 
is that the first and second fundamental forms of the boundary hypersurface 
$r=s$ shall be the same when calculated in both regions~\cite{Darmois}; this 
requires, in addition to continuity of the metric components at $r=s$, 
continuity of the first derivatives of $g_{zz}$, $g_{\phi\phi}$ and $g_{tt}$. 
We therefore have the conditions
\begin{gather}
\al^2(s)=A^2(s-k)^{8\si^2-4\si}, \qquad \be^2(s)=B^2(s-k)^{2-4\si},
         \qquad \ga^2(s)=C^2(s-k)^{4\si}, \label{metmatch} \\[10pt]
\frac{\al'}{\al}(s)=\frac{4\si^2-2\si}{s-k},  \qquad 
\frac{\be'}{\be}(s)=\frac{1-2\si}{s-k}, \qquad
\frac{\ga'}{\ga}(s)=\frac{2\si}{s-k}. \label{dermatch}
\end{gather}
Equations~(\ref{metmatch}) may be taken as defining the constants $A$, $B$ and 
$C$; in (\ref{dermatch}) there are three conditions for the two constants $\si$
and $k$, but since $\frac{\al'}{\al}(s)$, $\frac{\be'}{\be}(s)$ and 
$\frac{\ga'}{\ga}(s)$ satisfy (\ref{fe1}) with $p=0$, only two of these are
independent so they can always be satisfied. The simplest way to solve 
(\ref{dermatch}) is to focus on the second and third equations: adding them
eliminates $\si$ and so $k$ is found; $\si$ is then obtained from 
the third equation.  
\subsection{Perfect-fluid cylinder}
A cylindrically symmetric solution with a boundary arises from the choice
\begin{equation}   \label{cylbega}
\be\ga=r+ar^3-a^2br^5, \qquad a,b\ \text{constants},
\end{equation}
which satisfies (\ref{begacons}) and hence generates a cylindrically symmetric
solution. Eqn~(\ref{ga}) gives
\begin{equation}   \label{cylga}
\frac{\ga'}{\ga}=\pm\frac{ar\sqrt{6+5b-17abr^2+15a^2b^2r^4}}{1+ar^2-a^2br^4}.
\end{equation}
As to the choice of sign here, we continue initially with the positive sign as 
this leads to a positive density for the fluid. We then get from (\ref{be}) and
(\ref{al})
\begin{gather}
\frac{\be'}{\be}=\frac{1+3ar-5a^2br^4}{1+ar^2-a^2br^4}-\frac{\ga'}{\ga},  
                                                    \label{cylbe}  \\[10pt]
\frac{\al'}{\al}=\frac{\be'}{\be}-\frac{1}{\be\ga}.    \label{cylal} 
\end{gather} 
It is possible to integrate (\ref{cylga}) analytically and the other metric 
components are found from (\ref{cylbega}) and (\ref{cylal}); the results are
\footnote{We choose the constants of integration $L$ and $M$ so that the line 
element approaches the form \[ ds^2=dr^2+dz^2+r^2d\phi^2-dt^2 \] as $r\to 0$.}
\begin{equation}   \label{cylsoln}
\left.
\begin{split}
\ga=& L\left(-17b+30ab^2r^2+2b\sqrt{15}f(ar^2)\right)^{-\sqrt{15}/2}
                                                                \\[10pt]
&\times\left(\frac{12+10b-17bn+(-17b+30b^2n)ar^2+2f(n)f(ar^2)}
                          {ar^2-n}\right)^{-\frac{f(n)}{2\sqrt{1+4b}}} \\[10pt]
&\times\left(\frac{12+10b -17bm +(-17b+30b^2m)ar^2+2f(m)f(ar^2)}
                          {m-ar^2}\right)^{\frac{f(m)}{2\sqrt{1+4b}}}  \\[10pt]
 &\qquad \qquad \be = \ga^{-1}r(1+ar^2-a^2br^4),   \\[10pt]
 \qquad \qquad & \al= M \ga^{-1}(1+ar^2-a^2br^4)^{5/4}
          \left(\frac{m-ar^2}{ar^2-n}\right)^{\frac{2b-1}{4\sqrt{1+4b}}}, 
\end{split}
\right\}
\end{equation}
where
\begin{gather*}
m=\frac{1+\sqrt{1+4b}}{2b}, \qquad n=\frac{1-\sqrt{1+4b}}{2b},   \\[10pt]
f(x)=\sqrt{6+5b-17bx+15b^2x^2}.
\end{gather*}
The pressure and density of the fluid are
\begin{gather}
p=\frac{1}{8\pi\al^2}\left(\frac{3a-10a^2br^2}{1+ar^2-a^2br^4}\right), 
                                                     \label{cylpress} \\[10pt]
\rho=\frac{1}{8\pi\al^2}
\left(\frac{24a+20ab-102a^2br^2+120a^3b^2r^4+(30a^2br^2-9a)f(ar^2)}
       {(1+ar^2-a^2br^4)f(ar^2)}\right).   \label{cylden}
\end{gather}

A physically reasonable solution with a boundary is obtained by taking the 
constants $a$ and $b$ to be positive; in this case the pressure falls to zero
at $r=s=\sqrt{3/10ab}$. There are a few possibilities of misbehaviour which
must be checked: in the range $0\leq r\leq s$ and with $a$ and $b$ positive 
(i) the expressions $ar^2-n$ and $m-ar^2$ are positive since $n<0$ and
$m>as^2$ (ii) the polynomial appearing in $f(r)$ is positive since it
is positive at $r=0$ and has roots
\[
\pm\frac{1}{\sqrt{30ab}}\sqrt{17\pm\sqrt{-71-300b}}
\]
which are all complex (iii) the polynomial $1+ar^2-a^2br^4$ is also positive 
since it is positive at $r=0$ and the only one of its roots
\[
\pm\frac{1}{\sqrt{2ab}}\sqrt{1\pm\sqrt{1+4b}}
\]  
that is real and positive is greater than $s$.

Although the metric components (\ref{cylsoln}) are unwieldy to say the least, 
the quantities $\frac{\al'}{\al}$, $\frac{\be'}{\be}$ and $\frac{\ga'}{\ga}$ 
are quite manageable; in particular the match (\ref{dermatch}) to the exterior 
yields the simple relations
\begin{equation}   \label{cylmatch}
\si=\frac{3}{2\sqrt{9+20b}}, \qquad k=\frac{12}{45+100b}\sqrt{\frac{6}{5ab}}.
\end{equation}
In the exterior metric, which is valid for $r>s$, we must have $r-k>0$ to avoid
a singularity; this is indeed the case because
\begin{equation}   \label{cyls-k}
s-k=\sqrt{\frac{3}{10ab}}\left(\frac{21+100b}{45+100b}\right)
\end{equation}
is positive. 

As $b\to\infty$, $s\to 0$ so the cylinder disappears, and $\si\to 0$; the 
exterior metric approaches flat space--time in cylindrical coordinates. As 
$b\to 0$, $s\to\infty$ and we have a space-filling perfect fluid without a 
boundary; in this limit $\si\to\frac{1}{2}$ so the Levi--Civita metric can be 
matched to the cylinder for values of $\si$ up to, but not including, 
$\frac{1}{2}$. If the minus sign is taken in (\ref{cylga}) then the resulting 
solution still has pressure given by (\ref{cylpress}), so $s$ is unchanged, but
the density is now
\[
\rho=\frac{1}{8\pi\al^2}
\left(\frac{-24a-20ab+102a^2br^2+120a^3b^2r^4+(30a^2br^2-9a)f(ar^2)}
       {(1+ar^2-a^2br^4)f(ar^2)}\right),
\]
which (for positive $a$ and $b$) is negative in the range $0\leq r\leq s$; 
also, in the match (\ref{dermatch}) $k$ is unchanged from (\ref{cylmatch}) but 
$\si\to-\si$:
\begin{equation}   \label{cylmatch2}
\si=-\frac{3}{2\sqrt{9+20b}}, \qquad k=\frac{12}{45+100b}\sqrt{\frac{6}{5ab}}.
\end{equation}
Hence for this negative-mass solution the exterior metric may have $\si$ in the
range $0$ (where the interior vanishes) down to, but not including, 
$-\frac{1}{2}$. We have then a cylinder source for the Levi--Civita metric for 
the range
\begin{equation}      \label{cylrange}
-\sfrac{1}{2}<\si<\sfrac{1}{2},
\end{equation}
where we include the `no source' case $\si=0$. As is clear from 
(\ref{accel}) the cylinder is attractive for positive values of $\si$, 
which correspond to a positive density for the fluid, and repulsive for 
negative values of $\si$, corresponding to a negative density.

These results support the interpretation of the Levi--Civita metric in the 
range (\ref{cylrange}) as the relativistic line-mass field. As we remarked in 
Sec.~1 the Newtonian limit is given by $\si\to 0$, with $\si$ approaching 
the Newtonian mass per unit length. As is well known, there is in general 
relativity no unambiguous measure of the mass-energy of a system which is not 
asymptotically flat in all spatial directions; this we will demonstrate by 
using the Tolman mass formula~\cite{Landau} to give another measure of the mass
per unit length. The formula is
\[
M=\int (T^\al_{\ \al}-T^4_{\ 4})\sqrt{-g}\,d^3x,
\]
where the integral is over all space. It is, of course, intended to apply to
finite systems (but see~\cite{Florides}); the intention here is not to suggest 
that the Tolman formula can provide a correct measure of the mass per unit
length of a cylinder, but simply to give another example of a calculation of 
this quantity besides the acceleration of a test particle (\ref{accel}) (which 
suggests $\si$ as the mass per unit length). From the Tolman formula we take as
a measure of the mass per unit length
\begin{align}
m_T &=\int_0^{2\pi}\!\!\int_0^1\!\!\int^s_0(T^\al_{\ \al}-T^4_{\ 4})
                    \sqrt{-g}\,dr\,dz\,d\phi,   \nonumber  \\[10pt]
    &= 2\pi\int^s_0(3p+\rho)\sqrt{-g}\,dr,  \label{Tolman}
\end{align}
which for our solution gives the simple relation
\begin{equation}           \label{cylTolman}
m_T=\pm\frac{3}{40b}\sqrt{9+20b},
\end{equation}
where the plus and minus signs correspond, respectively, to choosing plus or
minus in (\ref{cylga}). Comparison of (\ref{cylTolman}) with (\ref{cylmatch})
and (\ref{cylmatch2}) shows that for small $\si$ (large $b$) 
$m_T\approx\si$ since, in the limit as $\si\to 0$, i.e.\ $b\to\infty$, $\si$ 
and $m_T$ behave like
\[
\si\approx\pm\frac{3}{2\sqrt{20b}}, \qquad m_T\approx\pm\frac{3}{2\sqrt{20b}}.
\]
Thus, as indicators of the gravitational mass, both $\si$ and $m_T$ agree in 
the Newtonian limit; however in the extreme relativistic regime, occurring when
$b$ is small, $\si$ approaches $\pm\sfrac{1}{2}$ whereas $m_T$ goes to 
infinitely positive or negative values. In fact, far from the Newtonian limit,
there are no cogent physical reasons for taking either of these as a measure of
a putative `mass per unit length' since such an idea has no well-defined
meaning. 
\subsection{Perfect-fluid wall}
We now present a plane-homogeneous solution with boundary. The 
$\phi$-coordinate is now Cartesian and hence we are constructing an infinite
wall of perfect fluid which we shall join to the exterior (Levi--Civita) 
space--time. The solution follows from the choice
\begin{equation}    \label{wallbega}
\be\ga=1+ar+a^2r^2-a^4br^4, \qquad a,b\ \text{constants},
\end{equation}
which is suitable for a plane-homogeneous space--time.
\footnote{\label{ft:oneside} We need only consider the $r\geq 0$-half of the 
system as the other half is identical, i.e.\ we could replace $r$ by $|r|$ in 
what follows.} 
Equation (\ref{ga}) then gives
\begin{equation}    \label{wallga}
\frac{\ga'}{\ga}=\!\!\pm\frac{\sqrt{-ac+(3a^3-2a^2c)r+\!(3a^4+6a^4b)r^2
          +\!(4a^4bc-2a^5b)r^3-\!9a^6br^4+10a^8b^2r^6}}{1+ar+a^2r^2-a^4br^4}.
\end{equation}
As with the cylinder, the positive sign in (\ref{wallga}) leads to a positive 
fluid density and we first consider this case. From (\ref{be}) and (\ref{al}) 
we get 
\begin{gather}
\frac{\be'}{\be}=\frac{a+2a^2r-4a^4br^3}{1+ar+a^2r^2-a^4br^4}-\frac{\ga'}{\ga},
                                     \label{wallbe}  \\[10pt]
\frac{\al'}{\al}=\frac{\be'}{\be}-\frac{c}{\be\ga}.  \label{wallal}
\end{gather}
The constant $c$ is now arbitrary and is a parameter for the solution.
The pressure and density are found to be {\footnotesize
\begin{gather}   
p=\frac{1}{8\pi\al^2}\left(\frac{a^2-6a^2br^2}{1+ar+a^2r^2-a^4br^4}\right),
                                               \label{wallpress} \\[10pt]
\rho\! =\!\!\frac{1}{8\pi\al^2}\!\! 
     \left(\frac{3a^3\!-\!2a^2c+\!(6a^4+\!12a^4b)r
+\!(12a^4bc-\!6a^5b)r^2\!-\!36a^6br^3+\!60a^8b^2r^5\!+(18a^4br^2\!-\!3a^2)g(r)}
      {(1+ar+a^2r^2-\!a^4br^4)g(r)}\right)     \label{wallden}
\end{gather}  }
where
\[
g(r)\!=\!\!\sqrt{-ac+\!(3a^3-2a^2c)r+\!(3a^4+6a^4b)r^2+(4a^4bc-\!2a^5b)r^3
                                     -\!9a^6br^4+10a^8b^2r^6}.
\]

In order to have a positive pressure and a boundary we must have $a>0$ and 
$b>0$. At $r=0$ the density is
\[
\rho(0)=\frac{1}{8\pi\al^2}\left[-3a^2
               +\left(2a-\frac{3a^2}{c}\right)\sqrt{-(ac)}\right],
\]
so for $\rho$ to be real and finite on the axis we require $c<0$. 
\footnote{The plane-symmetric case $c=0$ is therefore unphysical.}
Thus we confine the ranges of $a$, $b$ and $c$ as follows:
\begin{equation}    \label{ranges}
a>0, \qquad b>0, \qquad c<0.
\end{equation}
From (\ref{wallpress}) we see that the boundary is at 
$r=s=\sqrt{\frac{1}{6a^2b}}$. The function on the right of (\ref{wallga}) 
cannot be integrated analytically, however plots of $\ga$ for values of the 
constants satisfying (\ref{ranges}) show that it is well-behaved in its range 
of validity $0\leq r\leq s$. The functions $\al$, $\be$, $p$ and $\rho$ are 
also well-behaved in the range $0\leq r\leq s$ and moreover $p$ (up to the 
boundary) and $\rho$ are positive.

The match (\ref{dermatch}) to the exterior at $r=s$ gives
\begin{gather}   
\si=\frac{\sqrt{3\left(8a^2/b+27a^2+12a^3\sqrt{\frac{6}{a^2b}}-27ac
                       -6a^2c\sqrt{\frac{6}{a^2b}}\right)}}
  {18a+4a^2\sqrt{\frac{6}{a^2b}}}, \label{wallsi}  \\[10pt]
k=\frac{3-36b}{36ab+8a^2b\sqrt{\frac{6}{a^2b}}}.  \label{wallk}
\end{gather}
Since these give 
\[
s-k=\frac{5+36b+6ab\sqrt{\frac{6}{a^2b}}}{36a+8a^2b\sqrt{\frac{6}{a^2b}}}>0
\]
the exterior metric is free from singularities. 

What range of $\si$ does this source produce? As either $b$ or $c$ goes to zero
$\si$ approaches $\frac{1}{2}$; in the limit $b\to\infty$, $\si$ approaches a
constant value, but as $c\to -\infty$, $\si$ increases without bound. Hence, 
from (\ref{ranges}), we can match to the Levi--Civita metric for all $\si$
greater than $\frac{1}{2}$.

We still have the option of taking the negative sign in (\ref{wallga}). This
leads to the same pressure but we now have the density  
{\footnotesize
\[
\rho\! =\!\!\frac{1}{8\pi\al^2}\!\! 
     \left(\!\!\frac{-3a^3\!+\!2a^2c-\!(6a^4+\!12a^4b)r
-\!(12a^4bc\!-\!6a^5b)r^2\!+\!36a^6br^3-\!60a^8b^2r^5\!+(18a^4br^2\!-\!3a^2)
 g(r)}
      {(1+ar+a^2r^2-a^4br^4)g(r)}\!\!\right)    
\] }
which is negative in the range $0\leq r\leq s$. The match to the exterior only
differs from (\ref{wallsi}) and (\ref{wallk}) in the sign of $\si$:
\begin{gather}  
\si=-\frac{\sqrt{3\left(8a^2/b+27a^2+12a^3\sqrt{\frac{6}{a^2b}}-27ac
                       -6a^2c\sqrt{\frac{6}{a^2b}}\right)}}
  {18a+4a^2\sqrt{\frac{6}{a^2b}}}, \label{wallsi2}  \\[10pt]
k=\frac{3-36b}{36ab+8a^2b\sqrt{\frac{6}{a^2b}}}.  \label{wallk2}
\end{gather}
With the restriction (\ref{ranges}) we have a well-behaved solution 
in the range $0\leq r\leq s$, and the exterior now has any value of $\si$ less
than $-\frac{1}{2}$. Thus we have a wall source for the Levi--Civita metric
for
\begin{equation}  \label{wallrange}
|\si|>\sfrac{1}{2}.
\end{equation}

The Tolman mass formula can again be used, this time to calculate a mass per
unit area for the source. For this we take 
\[
m_T=\int_0^1\!\!\int_0^1\!\!\int_0^s (T^\al_{\ \al}-T^4_{\ 4})\sqrt{-g}\,
                                                  dr\,dz\,d\phi,
\]
which gives
\[
m_T=\pm\frac{\sqrt{8a^2/b+27a^2+12a^3\sqrt{\frac{6}{a^2b}}-27ac
                       -6a^2c\sqrt{\frac{6}{a^2b}}}}{12\pi\sqrt{3}},
\]
where the plus and minus signs correspond to positive and negative $\si$ 
respectively. Unlike the density $\rho$, $m_T$ is well-behaved when $c=0$; as 
$c\to-\infty$, $m_T$, like $\si$, goes to~$\pm\infty$. 

\section{Discussion}
In the previous section we constructed a perfect-fluid cylinder with 
positive density and pressure and found that matching to the Levi--Civita 
exterior was possible for $0\leq\si<\frac{1}{2}$; this is the same range of 
$\si$ as Bonnor and Davidson~\cite{Bonnor92b} found for their cylinder source. 
In addition, we matched a negative-density, perfect-fluid cylinder to the 
Levi--Civita metric for which $\si$ could take values in the range 
$-\frac{1}{2}<\si\leq 0$. We also showed that the Levi--Civita space--time 
represents the exterior field of a plane mass: we constructed a perfect-fluid 
wall with positive density and pressure which matches to the Levi--Civita 
metric for $\si>\frac{1}{2}$, and also a negative-density, perfect-fluid wall 
for which this match gives $\si<-\frac{1}{2}$.

The previous work on static cylinder sources in conjunction with the results
obtained here leads us to suspect that a perfect-fluid cylinder source
for the Levi--Civita space--time does not exist outside the range 
$-\frac{1}{2}<\si<\frac{1}{2}$. This exterior field provides a relativistic 
analogue of the Newtonian line-mass field; it has a clear Newtonian limit, 
given by $\si\to 0$, wherein  $\si$ approaches the Newtonian mass per unit 
length. There is however no justification for calling $\si$ the mass per unit 
length far from this limit and hence concluding that there is an upper limit on
the mass per unit length of a relativistic perfect-fluid cylinder; such an idea
has no well-defined meaning in general relativity as we have illustrated by 
using the Tolman mass formula to calculate the quantity $m_T$ given by 
(\ref{cylTolman}), which has as much a claim to be the ``mass per unit length''
as $\si$ in that it gives the correct Newtonian limit, but which takes the 
range $-\infty<m_T<\infty$ for our cylinder sources. 

What about cylinder sources other than perfect-fluid? If we allow arbitrary 
energy--momentum tensors then there \emph{do} exist cylinder sources for the 
Levi--Civita space--time outside the range $-\frac{1}{2}<\si<\frac{1}{2}$; for 
example, the metric given by
\begin{equation}   \label{string}
\al=\ga=\left(1+\frac{1+as^2}{s^2}r^2\right)^{\frac{as^2}{1+as^2}}, \qquad
\be=\frac{r}{1+\frac{1+as^2}{s^2}r^2}, \qquad \text{$a,s$ constants},
\end{equation}
describes a static cylinder with  energy--momentum tensor
\begin{equation}   \label{stringem}
T^r_{\ r}=T^\phi_{\ \phi}=\frac{a}{2\pi\al^2}\frac{\be^2}{r^2}
                        \left(1-\frac{r^2}{s^2}\right), \qquad
T^t_{\ t}=T^z_{\ z}=\frac{1}{4\pi\al^2}\frac{\be^2}{r^2}
         \left(\frac{1+as^2}{s^2}r^2-3-2as^2\right).
\end{equation}
At $r=s$, $T^r_{\ r}=T^\phi_{\ \phi}=0$ and a correct match may be made to
the Levi--Civita metric (\ref{Levi}), giving $\si=1$, $k=-\frac{2}{as}$. For 
positive $a$ and $s$ the complete solution is well-behaved everywhere and the 
cylinder has positive density and positive radial and azimuthal pressures, 
with a longitudinal stress equal to the density. This last property, 
$T^t_{\ t}=T^z_{\ z}$ (familiar from cosmic string theory), represents an 
exotic relativistic situation and so although for $\si=1$ at least, one can 
find both a cylinder and a wall source, we are inclined to the view that any 
cylinder source valid for $\si$ outside  $-\frac{1}{2}<\si<\frac{1}{2}$ will be
composed of rather bizarre relativistic material, at least in comparison to a 
perfect fluid.

What are we to conclude regarding the Levi--Civita space--time as the field 
outside a plane mass? It is more difficult to find a firm basis for the 
analysis here because of the lack of a Newtonian limit. Such a limit could be
identified by considering the proper acceleration of a test particle initially 
at rest with respect to the wall; in the Newtonian limit this acceleration 
would approach a constant value throughout the exterior and this constant is 
then $2\pi G$ times the Newtonian mass per unit area. But the acceleration in
question is given by (\ref{accel})
and this is not a constant for any real value of $\si$. This result is not so
surprising when we consider that in this putative Newtonian limit test 
particles at rest anywhere in the coordinate system of metric (\ref{Levi}), 
that is at rest relative to the wall, would have to experience the same 
acceleration away from the wall (to counteract the uniform force directed 
towards the wall). But a coordinate frame in which all points experience the
same proper acceleration cannot remain rigid in the sense that the proper 
distances between spacelike-separated points must change with time; hence the 
components of the metric tensor in this coordinate frame must depend on the 
time coordinate and therefore the Levi--Civita metric cannot achieve this 
limit.

Nevertheless we have found that the Levi--Civita space--time can represent
the exterior field of a wall of perfect fluid for $|\si|>\frac{1}{2}$. As 
remarked above, the usual suspects for the exterior field of a plane mass are 
the Taub metric ($\si=-\frac{1}{2}$) and the flat accelerated metric 
($\si=\frac{1}{2}$). These two appear as the bounding values of the exterior 
fields of the perfect-fluid wall (and cylinder) sources we have constructed, 
in the cases of negative and positive mass density respectively. It therefore
appears from this work that if perfect-fluid sources for these two metrics 
exist they are not continuously related to perfect-fluid sources for 
neighbouring values of $\si$. At any rate, the only perfect-fluid walls with 
boundary the author has found 
using the procedure of Sec.~2 and which upon matching to the Levi--Civita 
metric produce a range of $\si$ including $-\frac{1}{2},\frac{1}{2}$ are such 
that $s-k<0$, that is the exterior space--time has a plane singularity. In fact
these solutions are valid for \emph{all} values of $\si$ (including $\si=0$, 
where however the source does not vanish). Although an attempt has been made
to give some meaning to this type of plane singularity~\cite{Gron}, in this 
case we regard only singularity-free solutions as being of physical interest.

If we again allow for more unorthodox energy--momentum tensors, wall sources 
for $\si=-\frac{1}{2},\frac{1}{2}$ can certainly be found. A wall source for 
the Taub metric ($\si=-\frac{1}{2}$) is given by
\begin{gather*}
\al=\be=e^{ar}, \qquad \ga=e^{-\frac{ar^2}{4s}}, 
             \qquad \text{$a,s$ constants}\\[10pt]
T^r_{\ r}=\frac{a^2}{8\pi\al^2}\left(1-\frac{r}{s}\right) \qquad
T^z_{\ z}=T^\phi_{\ \phi}=-\frac{a}{16\pi s\al^2}
         \left(1-\frac{ar^2}{2s}\right),  \qquad 
T^t_{\ t}=\frac{a^2}{8\pi\al^2},
\end{gather*}
which matches to the Levi--Civita metric at $r=s$ with $\si=-\frac{1}{2}$, 
$k=s-\frac{2}{a}$. If $a$ and $s$ are positive the solution is well-behaved
everywhere and the wall has negative density (which tallies with its repulsive
exterior field); the radial pressure is positive up to the boundary and 
$T^z_{\ z}=T^\phi_{\ \phi}$ may be negative in some regions and positive in
others, depending on the value of $a$. A wall source for the accelerated 
metric $\si=\frac{1}{2}$ is
\begin{gather*}
\al=\be=e^{ar\left(1-\frac{r}{2s}\right)}, \qquad \ga=1+\frac{r}{s},
\qquad \text{$a,s$ constants} \\[10pt]
T^r_{\ r}=\frac{a}{8\pi s^2\al^2\ga}\left(1-\frac{r}{s}\right)
\left(2s+as^2-ar^2\right), \qquad
T^z_{\ z}=T^\phi_{\ \phi}=-\frac{a}{8\pi s\al^2},  \\[10pt]
T^t_{\ t}=-\frac{a}{8\pi\al^2}
  \left[\frac{2}{s}-a\left(1-\frac{r}{s}\right)^2\right].
\end{gather*}
This matches to the Levi--Civita metric at $r=s$ with $\si=\frac{1}{2}$, 
$k=-s$. The space--time is well-behaved everywhere if $a$ and $s$ are positive
and moreover the density is positive throughout the interior if 
$0<a<\frac{2}{s}$; then $T^z_{\ z}=T^\phi_{\ \phi}<0$ and $T^r_{\ r}\geq 0$. 
These two sources are physically unappealing and are presented here simply to 
show that wall sources for $\si=-\frac{1}{2},\frac{1}{2}$ exist; there is 
nothing in this work to support the interpretation of either of these metrics 
as the general-relativistic plane-mass field.

There is a further oddity of the wall sources, to be seen in the variation of
the proper acceleration of a test particle (\ref{accel}) with $\si$. We 
first rewrite (\ref{accel}) in Gaussian normal coordinates 
$(\bar{r},z,\phi,t)$, wherein $g_{\bar{r}\bar{r}}=1$; the Levi--Civita metric
in these coordinates is 
\begin{equation}  \label{Levigauss}
ds^2=d\bar{r}^2+A^2(\bar{r}-\bar{k})^{\frac{8\si^2-4\si}{4\si^2-2\si+1}}dz^2
+B^2(\bar{r}-\bar{k})^{\frac{2-4\si}{4\si^2-2\si+1}}d\phi^2
-C^2(\bar{r}-\bar{k})^{\frac{4\si}{4\si^2-2\si+1}}dt^2,
\end{equation}
and (\ref{accel}) becomes
\begin{equation}  \label{accel2}
\ddot{\bar{r}}=-\frac{2\si}{4\si^2-2\si+1}\left(\frac{1}{\bar{r}-\bar{k}}
                                                      \right).
\end{equation}
This form is preferable to (\ref{accel}) because it isolates the
$\si$-dependence. As a result $\ddot{\bar{r}}$ as a function of $\si$ has the 
same form regardless of the value of $\bar{r}$; this form is shown in Figure~1.
 
\begin{figure}
\begin{center}
\vspace{35mm}
\epsfig{file=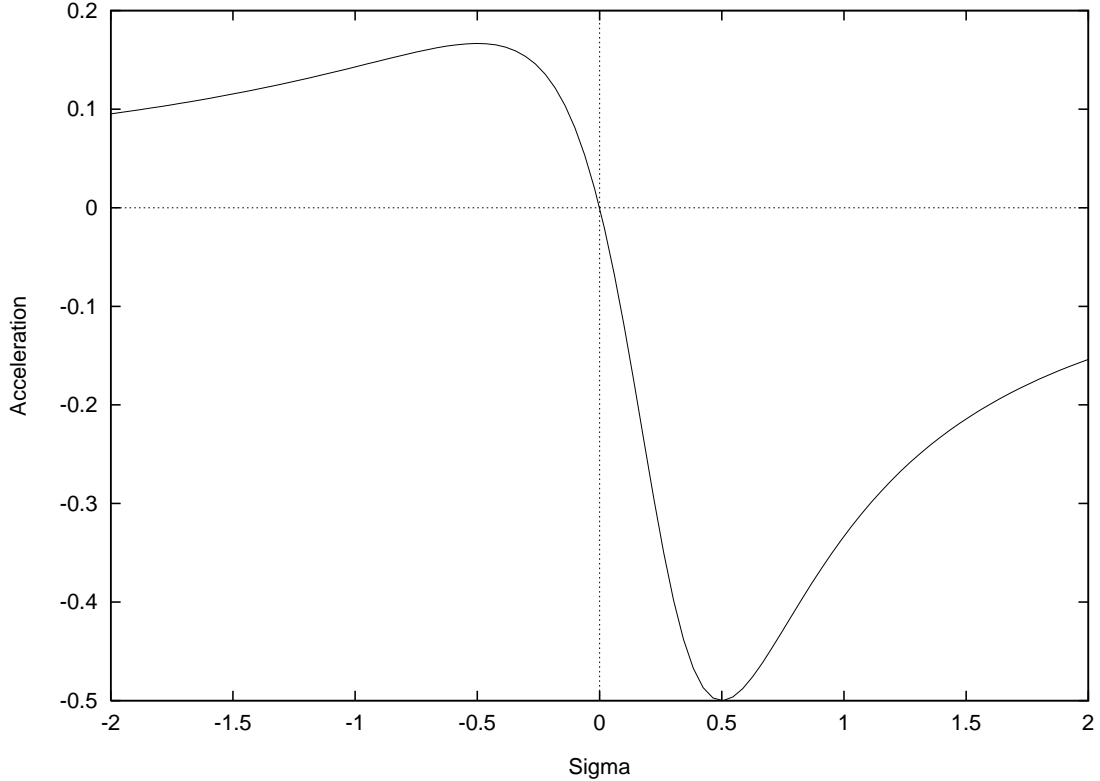, height=6cm}   \vspace{5mm}
\caption{Plot of $\ddot{\bar{r}}$ against $\si$ (see (\ref{accel2})) for 
$\bar{r}-\bar{k}=2$.}
\end{center}
\end{figure}

The proper acceleration increases as $\si$ goes from $0$ to $\frac{1}{2}$ ($0$ 
to $-\frac{1}{2}$) but falls to zero as $\si\to\infty$ ($\si\to-\infty$)!
Although the lack of a Newtonian limit means there is nothing we can term
the mass per unit area for even a limited range of $\si$, zero acceleration 
would be expected to occur only when the source vanishes or when a combination 
of negative density and positive pressure (or vice versa) conspires to 
produce zero gravitational mass. This is certainly not the case for the wall 
with positive density and pressure for which the limit $\si\to\infty$ is 
produced by $c\to-\infty$: the total gravitational mass cannot be zero and in 
no way does the limit $c\to-\infty$ correspond to the wall vanishing---the 
position of the boundary is unaffected and the density becomes infinite. 

Some insight into the nature of the system in this $\si\to\pm\infty$ limit is
afforded by the Levi--Civita line element in Gaussian normal coordinates
(\ref{Levigauss}). This form has a well-defined $\si\to\pm\infty$ limit which
is
\begin{equation}  \label{hyper}
ds^2=d\bar{r}^2+A^2(\bar{r}-\bar{k})^2dz^2+B^2d\phi^2-C^2dt^2.
\end{equation}
This metric is \emph{flat}, being transformable to a manifestly Minkowskian 
form by
\begin{equation}  \label{flattrans}
z'=(\bar{r}-\bar{k})\sin Az, \qquad r'= (\bar{r}-\bar{k})\cos Az.
\end{equation}
Metric (\ref{hyper}) has the general \emph{appearance} of flat space--time in 
cylindrical polar coordinates, with $z$ in the role of the angular coordinate,
and (\ref{flattrans}) that of the usual transformation from cylindrical polars 
to Cartesian coordinates. But $z$ is \emph{not} a periodic coordinate and if 
(\ref{flattrans}) were enforced globally it would have the effect of changing 
the topology of the exterior: (\ref{flattrans}) assumes that the points $z$ and
$z+2\pi/A$ are identified and consequently this transformation is only valid in
a local region covered by a range of $z$ smaller than $2\pi/A$. (Were we to 
take $z$ in the exterior as periodic we would have to do so in the interior 
also, thus producing a singularity in the source.) The geometry of the space 
part of (\ref{hyper}) is flat but rather bizarre: if we take two points on the 
boundary of the wall with different $z$-coordinate values and extend a straight
(we are in flat space) line perpendicular to the boundary from each point, then
the $z$-separation of these two straight lines is proportional to the distance 
from the wall, whereas the $\phi$-separation remains constant.

Although the $\si\to\pm\infty$ limit is not physically realizable, 
corresponding to a diverging density, and although we have nothing to call the 
mass per unit area, it is still curious that the acceleration in Figure~1 due
to the wall should approach zero as the density diverges. But, as discussed 
above, we are dealing here with a completely relativistic system; wall sources 
for the Levi--Civita space--time have no Newtonian limit to accommodate along 
with our Newtonian-based intuition about how a plane mass affects matter. An 
interesting question is: can a wall source be found in general relativity which
\emph{does} possess the Newtonian plane-mass limit? As remarked above, the 
exterior line element, in which the source is at rest, must depend on the time 
coordinate, so evidently this source will not be static. \vspace{10mm}

\noindent
{\bf \Large Acknowledgements} \\
This work was completed as part of an M.Sc.\ thesis and it is a pleasure to 
thank my supervisor Petros Florides for his guidance and advice. Thanks also to
Fabian Sievers and David Cashman for Figure~1.

\end{spacing}
\end{document}